\documentclass[preprints,article,accept,moreauthors,pdftex,10pt,a4paper]{mdpi}

\usepackage{amsmath}
\usepackage{bigints}
\usepackage{relsize}
\usepackage{slashed}
\usepackage[utf8]{inputenc}

\firstpage{1}
\pubvolume{xx}
\issuenum{1}
\articlenumber{5}
\pubyear{2019}
\copyrightyear{2019}
\history{Received: 28 February 2019; Accepted: 20 March 2019; Published: 22 March 2019}
\updates{yes} 

\Title{Effects of Hadron-Quark Phase Transitions in Hybrid Stars within the NJL Model}

\Author{Ignacio Francisco Ranea-Sandoval $^{1,2,}$*, Milva Gabriela Orsaria $^{1,2}$, Germ\'an Malfatti $^{1,2}$, Daniela Curin $^{1}$, Mauro Mariani $^{1,2}$, Gustavo An\'ibal Contrera $^{1,3,\dagger}$ and Octavio Miguel Guilera $^{4,5}$}
\AuthorNames{Ignacio Francisco Ranea-Sandoval, Milva Gabriela Orsaria, Germ\'an Malfatti, Daniela Curin, Mauro Mariani, Gustavo An\'ibal Contrera and Octavio Miguel Guilera}

\address{%
$^{1}$ \quad Grupo de Gravitaci\'on, Astrof\'isica y Cosmolog\'ia, Facultad de Ciencias Astron{\'o}micas y Geof{\'i}sicas, Universidad Nacional de La
  Plata, Paseo del Bosque S/N, 1900, Argentina; morsaria@fcaglp.unlp.edu.ar (M.G.O.); gmalfatti@fcaglp.unlp.edu.ar (G.M.); daniela.curin@ing.unlp.edu.ar (D.C.); mmariani@fcaglp.unlp.edu.ar (M.M.); contrera@fisica.unlp.edu.ar~(G.A.C.)\\
$^{2}$ \quad CONICET, Godoy Cruz 2290, 1425 Buenos Aires, Argentina\\
  $^{3}$ \quad IFLP, UNLP, CONICET, Facultad de Ciencias Exactas, Diagonal 113 entre 63 y 64, 1900 La Plata, Argentina\\
  $^{4}$ \quad Instituto de Astrof\'isica de La Plata, CONICET-UNLP, 1900, Argentina; oguilera@fcaglp.unlp.edu.ar\\
  $^{5}$ \quad Instituto de Astrof\'isica, Pontificia Universidad Catolica de Chile, 8970117, Chile
    }

\corres{Correspondence: iranea@fcaglp.unlp.edu.ar}

\firstnote{Current address: Department of Physics, San Diego State University, San Diego, CA 92182, USA.}

\abstract{We study local and non-local Polyakov Nambu-Jona-Lasinio models and analyze their respective phase transition diagram. We construct hybrid stars using the zero temperature limit of the local and non-local versions of Nambu-Jona-Lasinio model for quark matter and the modern GM1(L) parametrization of the non-linear relativistic mean field model for hadronic matter. We compare our models with data from PSR J1614-2230 and PSR J0343+0432 and also from GW170817 and its electromagnetic counterpart GRB170817A and AT2017gfo. We study observational signatures of the appearance of a mixed phase as a result of modeling a phase transition that mimics the Gibbs formalism and compare the results with the sharp first-order phase transition obtained using the Maxwell construction. We also study in detail the $g$-mode associated with discontinuities in the equation of state, and calculate non-radial oscillation modes using relativistic Cowling approximation.}


\begin{document}

\section{Introduction} \label{intro}
One of the purposes of Quantum Chromodynamics (QCD) is to describe how quarks and gluons interact to form the bound states composing hadronic matter, which can basically be classified into baryons and mesons. Experience indicates that under ordinary conditions of temperature and density, the phenomenon known as confinement occurs, in which quarks and gluons can only form such bound states. However, at sufficiently high densities and/or temperatures matter of unbound quarks and gluons, known by the generic name of quark-gluon plasma (QGP), can be formed. This has motivated the theoretical study of the possible phases of QCD and has revealed a complex phase structure. Thus, the study of the QCD phase diagram represents a theoretical and experimental challenge to date. The extreme conditions in which QGP is formed are difficult to produce under laboratory conditions, thus leaving few physical situations for experimental study, for example through heavy ion-collisions~\cite{doi:10.1146/annurev-nucl-101917-020852}. One of the natural scenarios where deconfined phases of QCD could appear is the interior of the so-called compact objects like neutron stars (NSs), which correspond to the region of high densities and low temperatures in the QCD phase diagram.

NSs are the densest known objects in the Universe with a mass, $M$$\sim$$1.4~M_\odot$ and a radius, $R$$\sim$$10~{\rm km}$. Astronomical observations of these objects can be used to test several high-energy physical theories~\citep{weber1999}. In the interior of NSs, matter is compressed several times the nuclear saturation density. Due such extreme conditions, it is expected the appearance of new degrees of freedom as hyperons and deconfined quark matter. However, the equation of state (EoS) describing the dense matter composing the cores of NSs is still unknown.

The observation of the 2 $M_{\odot}$ binary pulsars (PSR J1614-2230 \citep{Demorest:2010bx,Fonseca:2016tux} and PSR J0343+0432 \citep{Antoniadis:2013pzd,Lynch:2012vv}) has imposed a lower bound to the maximum mass of NSs. {{Therefore, a stiff enough EoS describing the dense matter in such objects should be able to fulfill this mass constraint.}} These mass measurements challenge high-density EoS and turn attention to the strange quark matter hypothesis{{, which suggests that the true ground state of matter is quark matter}} \citep{Bodmer:1971we,Witten:1984rs}.

{{On the other hand,}} the event of gravitational waves (GW) GW170817 emitted during a binary NS merger and its subsequent electromagnetic counterpart (GRB170817A and AT2017gfo) impose an upper limit to the effective tidal deformability, $\tilde{\Lambda}$, of the binary system. {{This parameter depends on the microphysical properties of the matter composition of the star and it is a measure of the influence of a star's internal structure on the waveform. Effects on this quantity related to the appearance of quark matter in the inner cores of compact stars have been studied in Refs.~\cite{2018PhRvD..97h4038P,2019PhRvD..99b3009C}.}}  Analysis of the data also serve to constraint the chirp mass of system, { {which determines (to leading-order) the evolution of binary NSs orbit and also the evolution of the GWs frequency signal during the inspiral phase}}. According the $\tilde{\Lambda} < 800$ constraint, several works conclude, to a $90\%$ level, that the radius of a $1.4~M_\odot$ NS, $R_{1.4}$, can not exceed $\sim$13.6 km (see, for example, Refs.~\cite{Annala:2017llu,2017ApJ...850L..34B,Fattoyev:2017jql,Raithel:2018ncd,Malik:2018zcf,2018PhRvC..98d5804T,jpg2019}). {{This is a clear example of the powerful tool that represents the detection of GWs for NS astrophysics, setting additional constraints on the EOS of dense matter composing such stars.}} It is expected that during LIGO's third observing run (planned to begin in April 2019) more NS-NS mergers would be detected from which new updates of the high-density EoS constraints can be established.

The possible existence of quark matter inside of NSs remains an open issue because QCD cannot be solved for dense matter in astrophysical environments with the tools usually employed by relativistic quantum field theories. Due the severe technical difficulties related to the non-perturbative nature of QCD at low and intermediate densities, several effective models have been proposed (MIT Bag model~\cite{Chodos1974}, Nambu-Jona-Lasinio (NJL) model~\cite{Nambu1, Nambu2} non-linear sigma model~\cite{PhysRevC.82.035803}, Field Correlator Method~\cite{SIMONOV1988, Mariani:2016pcx}). These models incorporate some (but not all) of the most relevant features of QCD.

The possibility of a hadron-quark matter phase transition in the inner core of NSs has been studied in several works (see, for example, Refs.~\cite{Bona:2012,Orsaria:2013,Orsaria:2014}) and some of the astrophysical implications of the existence of such dense matter have also been studied. In this paper we will work with both, the standard NJL and a non-local extension of the NJL model for the description of quark matter. We analyze the possibility of a hadron-quark phase transition in the interior of cold NSs using a modified GM1 relativistic mean field model, the modern GM1(L) parametrization, to describe the hadronic phase. {{In Refs.~\cite{Logoteta2012PhRvD..85b3003L,Logoteta2013PhRvC..88e5802L} the authors performed similar calculations constructing hybrid stars with quark matter cores and focusing on the finite-size effects, i.e, a geometric structure of the mixed phase or pasta phase structures with non-vanishing surface tension, which might be relevant in the hadron-quark interface. Also the authors of Ref.~\cite{Spinella:2018bdq} investigate the effect that a crystalline quark-hadron mixed phase can have on the neutrino emissivity from the cores of neutron stars considering the formation of spherical blob, rod, and slab rare phase geometries in the mixed phase. In this work we simulate the mixed phase through a continuous interpolation between the hadronic and quark matter phases that mimics the mixing or percolation. This interpolation function simulates the Gibbs construction and it will allow us to compare the results with the obtained considering a sharp phase transition through Maxwell construction.}}

The paper is organized as follows. In Section \ref{njl} we present the NJL models at finite temperature, with the inclusion of the Polyakov loop, and the zero limit equations to model the quark phase of cold NSs. We also present the phase diagram for each model. In Section \ref{astro} we analyze some astrophysical applications of the models focusing on the hybrid star EoS. We also describe the hadronic model used to study the hadron-quark phase transition, and we discuss briefly generalities for the occurrence of such transition. In addition, we present the formalism to study isolated NSs oscillation modes and calculate some frequencies, paying especial attention to the $g$-modes. Moreover we re-examine the {\it universal} relationship linked to the Constant Speed of Sound (CSS) parametrization obtained in Ref.~\cite{Ranea-Sandoval:2018bgu}. Finally, Section \ref{res} is devoted to summarize the obtained results and discuss their astrophysical relevance.

\section{Quark Matter Description within NJL Models} \label{njl}
In its simplest form, the Lagrangian of the NJL model contains local terms of scalar-isoscalar and pseudoscalar-isovectorial interaction between quarks, which are those that reproduce the chiral symmetry breaking dynamics of QCD. The development of this model was done without taking into account the phenomenon of confinement, since it was originally constructed to describe properties of nucleons, which do not exhibit this characteristic. Therefore, the absence of confinement is an important limitation of the theory. However, the lack of confinement can be partially remedied by including a dynamic variable known as Polyakov loop, which is treated as a background field that accounts for gluonic degrees of freedom that can reproduce the deconfinement transition. There are several anzats in the literature for the effective
Polyakov-loop potential, which accounts for gauge field
self-interactions~\cite{Andersen:2013swa}. In this work we use~\cite{Roessner:2006xn}
\begin{eqnarray}
{\cal{U}}(\Phi ,T) &=& \left[-\,\frac{1}{2}\, a_1(T, T_0)\,\Phi^2 + a_2(T, T_0)\, \ln(1 - 6\, \Phi^2 + 8\, \Phi^3 - 3\, \Phi^4)\right] T^4 \ ,\nonumber
\label{effpot}
\end{eqnarray}
where the coefficients $a_1(T, T_0)$ and $a_2(T, T_0)$ are fixed taking into account Lattice QCD simulations of gluon dynamics and the traced Polyakov loop $\Phi = [ 2 \cos(\phi_3/T) + 1]/3$ \cite{Contrera:2010,Rossner:2007}. The quantity $\phi_3$ is related with the color background field, as described in Section \ref{local}. The effective potential is key to describe the phase transition from the color confined state ($T < T_0$, with the minimum of the effective potential being at $\Phi = 1$) to the color deconfined state ($T > T_0$, with the minima of the effective potential at $\Phi = 0$), where $T_0$ is the critical temperature of the deconfinement phase transition. This parameter is the only free parameter of the Polyakov loop once the effective potential is fixed. Therefore, NJL-like models supplemented with the Polyakov loop accounts comfortably for chiral symmetry as well as for the deconfinement-confinement transition in QCD.

In addition to the confinement problem, the NJL model is non-renormalizable. Thus, it is necessary to apply some form of ultraviolet regularization with the introduction of a finite cutoff parameter. Many regularization procedures have been discussed in the literature, such as three and four-momentum cutoffs, Pauli-Villars regularization, proper-time regularization~\cite{Klevansky1992}. However, the introduction of a finite cutoff can be problematic to regularize quark loops integrals preserving some physics amplitudes in the meson sector. The way solve the drawbacks of the local NJL is to use non-local rather local interactions. Non-locality arises from several successful approaches to low-energy quark dynamics, such as one-gluon exchange descriptions~\cite{Ripka:1997zb,GomezDumm:2006vz,Contrera:2007wu}, the instanton liquid model~\cite{Diakonov:1985eg,Schafer:1996wv}, and~the Schwinger-Dyson resummation technique~\cite{Roberts:1994dr,Roberts:2000aa}. The non-local extensions of the NJL model are designed to remove the deficiencies of the local theory.

The general expression for the mean-field thermodynamic potential in NJL-like models can be written as
\begin{equation}
\Omega = \Omega^{\rm reg} + \Omega^{\rm free} + \Omega_0
+\mathcal{U}(\Phi ,T)\, ,
\label{eq:Omega}
\end{equation}
where $\Omega_0$ is defined by the condition that $\Omega$ vanishes at
$T=\mu=0$.

Once the thermodynamical grand potential is determined, the system's pressure $P=-\Omega$, quark number density $n_q = \sum_f n_f,$ where $f$ runs over all quark flavors. With these quantities, the EoS for the quark matter $\epsilon (P) = - P + T S + \sum_f \mu_f n_f$ (where $S = \frac{\partial P}{\partial T}$ and $n_f = \frac{\partial P}{\partial \mu_f}$), can be also~calculated. 

\subsection{Local SU(3) PNJL Model}
\label{local}

The regularized and free thermodynamic potentials for the SU(3) local NJL model at finite temperature are given by
\begin{eqnarray}
\Omega^{\rm reg} = &-& 2
N_c\sum_{f}\int_0^{\Lambda}\frac{\mathrm{d}^3p}
{\left(2\pi\right)^3}\,{E_f} + G_s (\alpha^2 + \beta^2 + \gamma^2) + 4
H \alpha \beta \gamma  \nonumber \\
\Omega^{\rm free} = &-&2T \sum_{f, c}\int
_0^{\infty}\,\frac{\mathrm{d}^3p}
{\left(2\pi\right)^3}\,\Bigg\{ {\mathrm{ln} \left(1+ e^{-\frac{E_f -
    \mu_f + i c\phi_3}{T}}\right)} +
{\mathrm{ln} \left(1+ e^{-\frac{E_f + \mu_f - i
    c\phi_3}{T}}\right) \Bigg\} }
\, , \label{omegaT_njl}
\end{eqnarray}
where, $\alpha = \left\langle\bar{\psi_{u}}\psi_{u}\right\rangle$, $\beta = \left\langle\bar{\psi_{d}}\psi_{d}\right\rangle$ and $\gamma = \left\langle\bar{\psi_{s}}\psi_{s}\right\rangle$ are the condensates corresponding to each quark flavor. Since we are consider three quark colors, $N_c = 3$ and  $E_{f} = \sqrt{p^{2}+M_{f}^{2}}$. The color background fields due the coupling to the Polyakov loop are $ \phi_c = c \phi_3 = n_3 \phi_3$, i.e., $\phi_r = - \phi_g = \phi_3$ and $\phi_b = 0$. The sums over flavor and color indices run over $f = (u, d, s)$
and $c = ( r, g, b)$, respectively. The constituent quark masses $M_{f}$ are given by
\begin{equation}
M_{f}=m_{f}-2G_s\langle\bar{\psi_{f}}\psi_{f}\rangle - 2H
\langle\bar{\psi_{j}} \psi_{j} \rangle\langle\bar
                {\psi_{k}} \psi_{k} \rangle \, , \nonumber
\end{equation}
with $f,j,k=u,d,s$ indicate cyclic permutations. 

The scalar coupling constant, $G_s$, the 't Hooft coupling constant, $H$, quark masses and the three-momentum ultraviolet cutoff, $\Lambda$, are model parameters. Their values, $m_u=m_d=5.5$ MeV, $m_s=140.7$ MeV, $\Lambda=602.3$ MeV, $G_s\Lambda^{2}=3.67$ and $H\Lambda^{5}=-12.36$, are taken from Ref.~\cite{Rehberg1996}. 

The minimization of the thermodynamic potential with respect to the quark condensates and the Polyakov loop color field $\phi_3$ leads to a system of coupled non-linear equations that can be solved~numerically. 

The solution of the system equations allow us to construct the phase diagram of the model and to get the thermodynamical quantities to determine the quark matter EoS.

\subsubsection*{Zero Temperature Limit Including Vector Interaction}

At zero temperature, the Polyakov loop vanishes. {{In this limit, we include a repulsive vector interaction among quarks. The repulsive character of the vector coupling in NJL-like models shifts chiral restoration to larger values of the quark chemical potential in the phase diagram~\cite{Fukushima:2008wg,Fukushima:2008b}. This fact affects the quark-hadron phase transition. Therefore, if the quark deconfined transition in the cores of NSs is modeled through NJL-like models, it is expected that the vector coupling contribution modifies the hybrid EoS and hence the mass-radius relationship of NSs. Vector interactions are crucial to stiffen the quark EoS for the astrophysical application of the NJL model.}} The regularized thermodynamic grand potential is the same of Equation~(\ref{omegaT_njl}). The free thermodynamic potential is modified at zero~temperature
\begin{eqnarray}
\Omega ^{\rm free} (T \rightarrow 0) &=&   -
\sum_{f=u,d,s}\int
_0^{p_{F_f}}\,\mathrm{d}{p}\,~\frac{p^4}{E_f} 
\, , \label{omega_njl}
\end{eqnarray}
where $p_{F_f} = \sqrt{\mu_f^2-M_{f}^2}$. The contribution to the grand potential due vector interaction is given by
\begin{equation}
\Omega^{\mbox{v}} = - G_{\mbox{v}}\,\sum_f n_f^2
\end{equation}
where the quark number density of flavor $f$ in the mean field approximation is given by
\begin{equation}
n_f=\frac{N_c}{3\pi^2}[(\mu_f - 2G_{\mbox{v}} n_f)^2-M_f^2]^{3/2} \, .
\end{equation}
The vector coupling
constant $G_{\mbox{v}}$ is treated as a free parameter and it is usually described in terms of the strong coupling constant $G_s$. From here on, we will use $\xi_{\mbox{v}}^{L} = G_v/G_s$. Vector
interaction shifts the quark chemical potential by
\begin{equation}
\mu_f\, \rightarrow \mu_f - 2\xi_{\mbox{v}}^{L} G_s n_f.
\end{equation}
The system of non-linear equations at zero temperature are solved together with the condition of charge neutrality and baryonic number conservation to obtain the EoS which can be applied to describe quark matter in the interior of NSs. 

\subsection{Non-Local SU(3) Model}

The quantities $\Omega^{\rm reg}$ and $\Omega^{\rm free}$ are given by
\begin{eqnarray}
\Omega^{\rm reg} = &-&4 T \sum_{f,c}\sum_{n=0}^{\infty}\left[\int\frac{d^3p}{(2\pi)^3}
 \, \log \left[\frac{w_{fnc}^2 + M_{f}^2(w_{fnc}^2)}
   {w_{fnc}^2 + m_f^2}\right]\right] -\frac{1}{2} \left( \sum_{f}\left(\bar\sigma_f \bar S_f + \frac{G_s}{2}\bar S_f^2\right) +
   \frac{H}{2}\bar S_u \bar S_d \bar S_s\right) \nonumber
\\ \Omega^{\rm free} = &-&2T \sum_{f,c}\int\frac{d^3p
 }{(2\pi)^3}\left[{\rm ln} \left(1+ e^{-\frac{E_f -
    \mu_f - i c\phi_3}{T}}\right) +{\rm ln} \left(1+ e^{-\frac{E_f + \mu_f + i
    c\phi_3}{T}}\right)\right] \, ,
\end{eqnarray}
where $E_f=\sqrt{\vec p^{\;2} + m_f^2}$, $w_{fnc}^2 = (w_n -
  i\mu_f + \phi_c)^2 + \vec p^{\,2}$, $w_n$ denote Matsubara
  frequencies. $\bar S_f$ are auxiliary fields obtained by minimizing the thermodynamic potential, $\Omega$, with respect to
the mean field values $\bar \sigma_f$ \cite{Scarpettini:2003fj,Contrera:2007wu}.  Similar to the local case, these equations plus minimizing
$\Omega$ with respect to the Polyakov loop color field $\phi_3$ leads to a system of coupled non-linear equations that can be solved
numerically to obtain the mean field values $\bar \sigma_f$ and the traced Polyakov loop. The momentum dependent constituent
quark masses are given by
\begin{equation}
M_{f}(w_{fnc}^2) \ = \ m_f\, + \,
\bar\sigma_f\, R(w_{fnc}^2).
\end{equation} 

To regulate the non-local interactions we use a Gaussian form factor
$R(w_{fnc}^2) = e^{-{w_{fnc}^2}/{\Lambda^2}}$, where $\Lambda$ is
relevant for the stiffness of the chiral transition. The up $m_u$ and
down $m_d$ current quarks masses and the coupling constants
$G_s$, $H$, and $\Lambda$ are chosen so as to reproduce the
phenomenological values of the pion decay constant $f_\pi=92.4$ MeV and the
meson masses $m_{\pi}=139.0$ MeV, $m_K=495.0$ MeV, $m_{\eta'}=958.0$ MeV~\cite{Contrera:2007wu, Contrera:2009hk}, leading to $m_s = 127.77$ MeV, $\Lambda = 780.63$ MeV, $G_s \Lambda^2 = 14.48$, and $H \Lambda^5 = -267.24$. Light quark current masses are set to $ m_u = m_d = 5.50$ MeV.

\subsubsection*{Zero Temperature Limit Including Vector Interaction}
{{As in the local NJL model, at zero temperature we include vector interaction to get a stiffer non-local quark matter EoS for the construction of hybrid stars.}} The mean-field regularized and free thermodynamic potentials for the SU(3) non-local NJL model can be expressed as
\begin{eqnarray}
\Omega ^{\rm reg}  (T \rightarrow 0) = &-&\frac{N_c}{\pi^3}\sum_{f=u,d,s}
  \int^{\infty}_{0} dp_0 \int^{\infty}_{0}\, dp \, p^2 \,\mbox{ ln }\left[
  \frac{ \widehat{\omega}_f^2 + M_{f}^2(\omega_f^2)}{\omega_f^2 + m_{f}^2}\right] \nonumber \\
  &-&\frac{1}{2}\left[ \sum_{f=u,d,s} (\bar \sigma_f \ \bar S_f +
    \frac{G_S}{2} \ \bar S_f^2) + \frac{H}{2} \bar S_u\ \bar S_d\ \bar
    S_s\right]\\ \nonumber
   \Omega ^{\rm free}   (T \rightarrow 0) = &-&
  \frac{N_c}{\pi^2} \sum_{f=u,d,s} \int^{\sqrt{\mu_f^2-m_{f}^2}}_{0}
  dp\,\, p^2\,\, \left[(\mu_f-E_f) \Theta(\mu_f-m_f) \right]
\end{eqnarray}
where $N_c=3$, $\Theta$ is the Heaviside function, $\omega_f^2 =(\,p_0\,+\,i\,\mu_f\,)^2\,+\,p^2$ and $E_{f} =\sqrt{p^{2} + m_{f}^{2}}$.  The constituent quark masses, $M_{f}$, are treated as momentum-dependent quantities expressed as
\begin{equation}
M_{f}(\omega_{f}^2) \ = \ m_f + \bar\sigma_f R(\omega_{f}^2)\, \nonumber,
\end{equation}
where $R(\omega^2_f)$ denotes the Gaussian form factor we use, $R(\omega^2_f) = \exp{\left(-\omega^2_f/\Lambda^2\right)}$ (for results using different form factors, see for example Ref.~\cite{Aguilera2006}) and the quantity $\widehat{\omega}$  is affected by the vector interaction contribution~\cite{Orsaria:2014}, which shifts the quark chemical potential
\begin{equation}
\mu_f\, \rightarrow \widehat{\mu}_f=\mu_f - R(\omega^2_f)\bar\theta_f\, ,
\label{shift}
\end{equation}
where $\bar\theta_f$ denotes the vector mean fields.
The contribution to the grand potential due vector interaction is given by
\begin{equation}
\Omega^{\mbox{v}} = - \,\sum_f \frac{\bar\theta_f^2}{4 \xi_{\mbox{v}}^{NL} G_s},
\end{equation}
where $\xi_{\mbox{v}}^{NL} = G_{\mbox{v}}/G_s$. 

Please note that the
quark chemical potential shift in Equation~(\ref{shift}) does not affect the non-local form factor, avoiding a recursive problem (see Ref.~\cite{Orsaria:2014} and references therein).

As in the local model the charge neutrality and  baryonic number conservation conditions, together with the minimization of the grand canonical potential with respect to the $\bar \sigma_f$ fields, determine the astrophysical quark matter EoS.

\subsection{QCD Phase Diagram}
\begin{figure}[H]
  \centering
  \includegraphics[width=0.6\textwidth,angle=0]{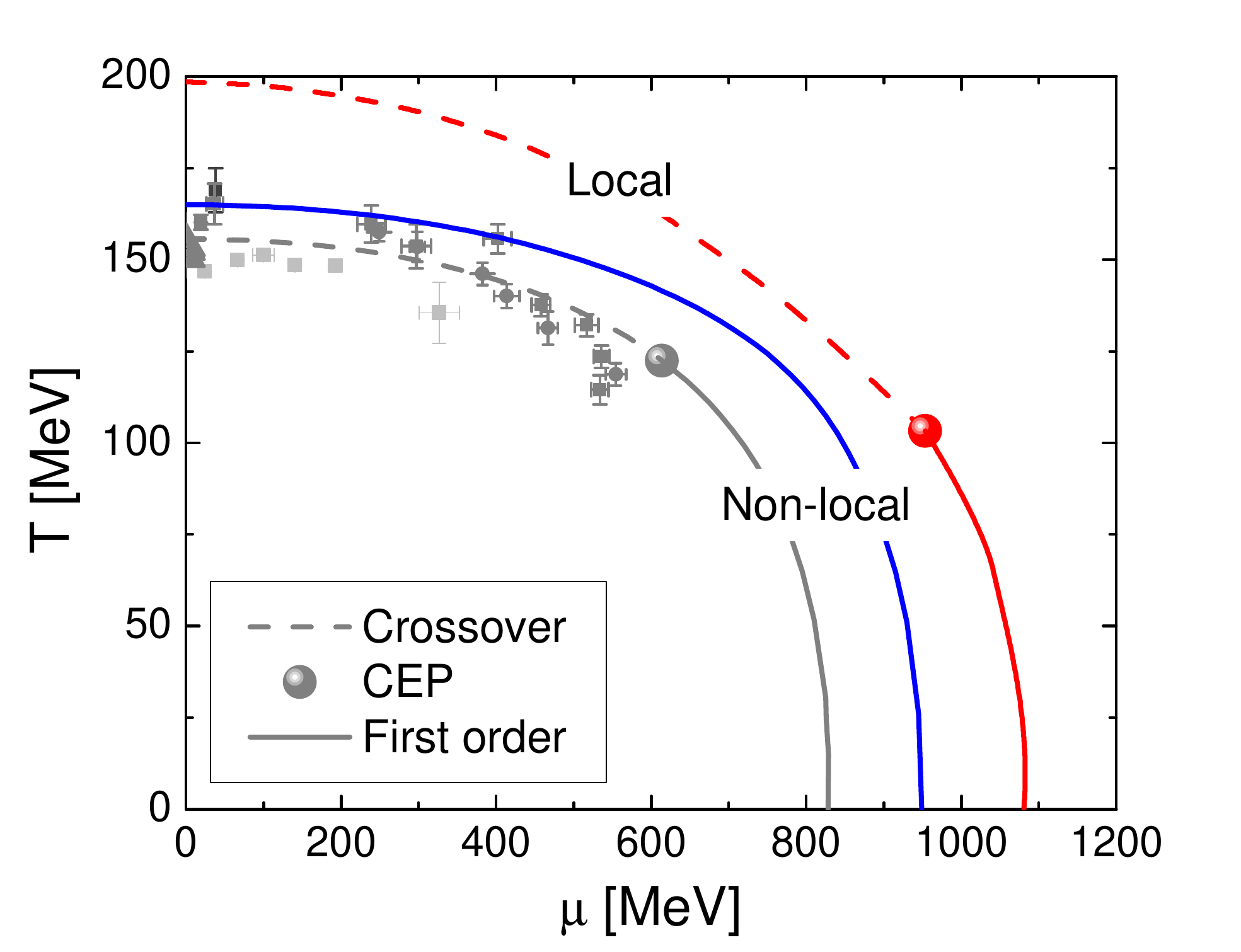}
  \caption{Phase diagram of the NJL models considered in this work. As a comparison, in gray we show the results obtained with a non-local NJL parametrization considering $m_s = 95$ MeV. The solid line corresponds to the non-local parametrization with $m_s = 127.77$ MeV.  The critical endpoint, CEP, of each phase equilibrium curve, crossover or first order, is denoted by a round dot. The different gray dots indicate some LQCD results and signals of deconfinement from the determination of the chemical freeze-out (see Ref.~\cite{jpg2019} and references therein).} 
  \label{qcd-diag}
\end{figure}

According the QCD phase diagram, it is known that the chiral symmetry is spontaneously broken at low temperature, but it is recovered above a certain value. The local order parameter, the quark condensate, is different from zero at low temperature, where the chiral symmetry is broken, and is zero above the chiral phase transition.  Lattice QCD simulations suggest that for fermions without mass, the chiral restoration and deconfinement take place at the same temperature, at least in the case of zero chemical potential~\cite{Ratti_2018}. In this case, the chiral symmetry restoration temperature is $T_c \simeq 155-205$ MeV, depending on the number of flavors. As a way to compare both the local and non-local models we have chosen parametrizations in which the values of the light quark condensates at zero chemical potential are comparable ($\alpha^{1/3} =\beta^{1/3} =- 241.9$ MeV for the local model and $\left\langle\bar{\psi_{u}}\psi_{u}\right\rangle^{1/3} = \left\langle\bar{\psi_{d}}\psi_{d}\right\rangle^{1/3} = - 243.8$~MeV for the non-local model). It is important to note that for moderate quark mass values, the chiral transition does not have a well-defined defined order parameter, and a pure phase transition does not occur but only a rapid change, called  {\it crossover}, present in the gray line and the labeled ``Local'' {lines of Figure} \ref{qcd-diag}. From ab-initio lattice QCD simulations follow that in the QCD phase diagram the transition on the temperature axis is a {\it crossover} at zero chemical potential~\cite{Ratti_2018}. 

The QCD phase transition occurring at high temperature and low densities is extremely relevant as the spectrum of primordial GWs changes dramatically if such phase transition is sharp or a smooth {\it crossover}. The main reason of these differences are based of the fact that if the primordial hadron-quark phase transition is sharp, bubble nucleation process produce larger perturbations which excite more efficiently the production of GWs~\cite{PhysRevD.83.064030}. The local NJL, show a {\it crossover} phase transition compatible with previous results obtained using different parametrizations  of the NJL-like models. The results obtained with the non-local version of the NJL model is a first order phase transition which is qualitatively similar with the result obtained with the standard MIT bag model~\cite{PhysRevD.83.064030}. In this way, data from primordial GWs could serve to probe the nature of the phase transition and also the existence or not of the QCD critical end point.

\section{Astrophysical Applications of NJL EoS} \label{astro}

High-energy astrophysics, especially X-ray and Radio astronomy, and more recently GW astronomy, can be used to shed some light into the nature of matter inside a NS (for details, see Ref.~\cite{jpg2019} and references therein). {{There are several works where the NJL model has been employed to describe a possible quark matter phase in NSs (see for example Refs.~\cite{LAWLEY200529, Contrera:2016phj, Ranea:2016}}.} {{Local and non-local extensions of the NJL model at zero temperature can be used to study hybrid stars with and/or without a quark-hadron mixed phases in their inner cores to explore the role of quark deconfinement in NSs.}}

\subsection{Hybrid Stars}
\label{hyb}
Matter in the inner core of NSs is compressed to densities several times higher than the saturation density, $\rho _0$. Under these conditions, a phase transition in which nuclei dissolve to form a QGP may occur. When pure QGP appears in the interior of a compact object we will call it a quark-hybrid star. Objects in which the QGP appears forming a mixed phase which coexist with hadronic matter will be denoted as hybrid stars.

In this work, we will construct hybrid and quark hybrid stars, using the NJL models described in the previous section {{to model the QGP phase in their inner core.}}

To describe the hadronic phase (outer core) of the hybrid stars we will use the GM1(L) extension of the classical GM1 parametrization in the Relativistic mean field theory (RMF), including both hyperons and Delta isobars~\cite{Spinella2017:thesis}.

\subsubsection{Hadronic EoS}
{\textls[-5]{To describe matter in the outer core of our hybrid star models, we will use the relativistic mean field theory taking into account the GM1(L) parametrization.}} The main change from the classical GM1 model to this improved version is the inclusion of a density-dependent isovector coupling constant for the meson-baryon interaction~\cite{Spinella2017:thesis}. The exponential coupling constant introduces an extra parameter that is fixed to fit the experimentally determined slope of the asymmetry energy, L, at saturation density. The mediators of the interaction are the mesons  $\sigma$ (scalar field), $\omega$ (vector field) and \textbf{$\rho$} (isovector field), with bare masses $m_\sigma$, $m_\omega$ and $m_rho$, respectively. The Lagrangian of this model read as
\begin{equation}
\begin{array}{lll}
  \mathcal{L} &=& \sum_{B}\bar{\psi}_B \bigl[\gamma_\mu [i\partial^\mu - g_{\omega B}  \omega^\mu
     - g_{\rho B}(n) {\boldsymbol{\tau}} \cdot {\boldsymbol{\rho}}^\mu] - [m_B - g_{\sigma B}\sigma]
    \bigr] \psi_B + \frac{1}{2} (\partial_\mu \sigma\partial^\mu  \sigma  - m_\sigma^2 \sigma^2) \\
  & -& \frac{1}{3} b_\sigma m_N [g_{\sigma N} \sigma]^3 - \frac{1}{4} c_\sigma [g_{\sigma N} \sigma]^4
  - \frac{1}{4}\omega_{\mu\nu} \omega^{\mu\nu} +\frac{1}{2}m_\omega^2\omega_\mu \omega^\mu + \frac{1}{2}m_\rho^2
  {\boldsymbol{\rho\,}}_\mu \cdot {\boldsymbol{\rho\,}}^\mu - \frac{1}{4}
  {\boldsymbol{\rho\,}}_{\mu\nu} \cdot {\boldsymbol{\rho\,}}^{\mu\nu} \,  \label{eq:Blag}
  \end{array}
\end{equation}
where $g_{\sigma B}$, $g_{\omega B}$ and $g_{\rho B}(n)$ are the meson-baryon coupling constants, being the last one density
dependent, where $n = \sum_B n_B$ is the (total) baryon number density. In this work we consider that the baryons $B$ include the spin $1/2$ baryon octet compounded by the nucleons $N=(n,p)$ and hyperons $Y=(\Lambda, \Sigma^-, \Sigma^0, \Sigma^+, \Xi^-, \Xi^0)$, as well as the spin $3/2$ delta isobar quartet $(\Delta^-, \Delta^0, \Delta^+$, $\Delta^{++})$. In addition to the interaction terms meson-baryon, two non-linear terms were included taking into account the meson-meson interaction for the sigma meson.

The density dependence of the \textbf{$\rho$} meson coupling constants is given by
\begin{equation}
g_{\rho B}(n) = g_{\rho B}(n_0)\,\mathrm{exp}\left[\,-a_{\rho} \left( \frac{n}{n_0} - 1\right)\,\right],
\end{equation}
 This choice of parametrization accounts for nuclear
medium effects by making the meson-baryon coupling constants dependent
on the local baryon number density~\cite{Fuchs:1995as}.  The
parameter $a_{\rho}$ is adjusted to satisfy the constraints on the asymmetry
energy slope at saturation density $L_0$ without affecting the other saturation
properties, leaving the same values for the rest of the GM1 parameters.

The baryon and meson field equations are obtained by evaluating the
Euler-Lagrange equations for the fields in Equation~(\ref{eq:Blag}). Then,
applying the RMF approximation to the system, the
meson mean-field equations are given by
\begin{eqnarray}
m_{\sigma}^2 \bar{\sigma} &= & \sum_{B} g_{\sigma B}\, n_B^s- b_{\sigma} \, m_N\,g_{\sigma N}(g_{\sigma N}\,\bar{\sigma})^2 -
 c_{\sigma} \, g_{\sigma N} \, (g_{\sigma N}\, \bar{\sigma})^3 ,\nonumber\\
  m_{\omega}^2 \bar{\omega} &=& \sum_{B} g_{\omega B} n_{B} , \\
m_{\rho}^2\bar{\rho}&=&  \sum_{B}g_{\rho B}(n)\,I_{3B}\, n_{B} \nonumber
\end{eqnarray}
where $\bar{\sigma}\equiv\langle\sigma\rangle$, $\bar{\omega}\equiv\langle\omega\rangle$, $\bar{\rho}\equiv\langle\rho\rangle$ and $I_{3B}$ is the 3-component of isospin. $n_{B}^s$ and $n_{B}$ are the scalar and particle number densities for each baryon $B$, respectively given by
\begin{eqnarray}
n_{B}^s&\equiv& \langle \bar\psi_B \psi_B\rangle = \frac{\gamma_B}{2\pi^2}  \bigintss_0^{k_B} \frac{m^*_B(\bar{\sigma})}{\sqrt{k^2 + m^{*2}_B(\bar{\sigma})}}\, k^2 dk = \mathlarger{\mathlarger{\sum}}_B \frac{\gamma_B}{2\pi^2} \left( \frac{m^{*2}_B}{2}\right) \left[ k_B E_B - m^{*2}_B \ln \left(\frac{E_B+k_B}{m^*_B}\right) \right],  \nonumber \\
n_{B}&\equiv& \langle \psi_B^\dagger \psi_B\rangle = \frac{\gamma_B}{2\pi^2} \bigintsss_0^{k_B} k^2 dk = \frac{\gamma_B}{2\pi^2} k_B^3,
\end{eqnarray}
where $\gamma_B=2J_B+1$ is the corresponding spin degeneration factor, $k_B$ is the Fermi momentum, $m_B^*= m_B - g_{\sigma B}(n)\bar{\sigma}$ is the effective mass and $E_B=\sqrt{k_B^2 + m^{*2}_B(\bar{\sigma})}$ is the Fermi energy.

Now, to reproduce the NS matter conditions, after including the leptons in the model as a Fermi free gas, electron charge and baryon number must be conserved, according respectively to
\begin{equation}
\sum_B q_B\,n_B + \sum_l q_l\,n_l = 0 \;\;\;\;\;\; \textrm{and} \;\;\;\;\;\; \sum_B n_B - n = 0,
\end{equation}
where the subscripts $q_B$ and $q_l$ corresponds to baryons and leptons electric charges, respectively, in units of the elementary charge.

This constitutes a system of five coupled non-linear equations that are solved simultaneously to determine the meson mean-fields ($\bar{\sigma}$, $\bar{\omega}$ and  $\bar{\rho}$) and the neutron and electron Fermi momenta ($k_n$ and $k_e$). The Fermi momenta of the rest of the baryons are governed by the condition that NS matter be in chemical equilibrium,

\begin{equation}
\mu_B = \mu_n - q_B\, \mu_e
\end{equation}
where $\mu_B$ is the baryon chemical potential given by
\begin{equation}
\mu_B = \sqrt{k_B^2 + m^{*2}_B(\bar{\sigma})} + g_{\rho B}(n)\, \bar{\rho}\, I_{3B} + \widetilde{R}.
\end{equation}
$\widetilde{R}$ is a rearrangement term,
\begin{equation}
\widetilde{R} =\sum_B  \frac{\partial g_{\rho B}(n)}{\partial n} I_{3B}\, n_B\, \bar{\rho} ,
\label{rear}
\end{equation}
due the density dependence of the coupling constants. The inclusion of Equation~(\ref{rear}) is mandatory for thermodynamical consistency~\cite{Hofmann:2001}.  The rearrangement term also affects the baryonic pressure,
\begin{equation}
\begin{array}{lll}
P &=& \mathlarger{\mathlarger{\sum}}_B \,\frac{\gamma_B}{6\pi^2} \bigintss_0^{k_B} \frac{k^4 \, dk}{\sqrt{k^2 + m^{*2}_B(\bar{\sigma})}} + \mathlarger{\mathlarger{\sum}}_l \, \frac{1}{3\pi^2} \bigintss_0^{k_l} \frac{k^4 \, dk}{\sqrt{k^2 + m^{2}_l(\bar{\sigma})}}\\
&-& \frac{1}{2} m_{\sigma}^2\, \bar{\sigma}^2 + \frac{1}{2} m_{\omega}^2 \,\bar{\omega}^2 + \frac{1}{2} m_{\rho}^2 \,\bar{\rho}^2
- \frac{1}{3} b_{\sigma} m_N (g_{\sigma N}\, \bar{\sigma})^3 - \frac{1}{4} c_{\sigma} (g_{\sigma N}\, \bar{\sigma})^4 + n \, \widetilde{R}\, ,
\end{array}
\end{equation}
but it is not present in the energy density,
\begin{equation}
\begin{array}{lll}
\epsilon &=& \mathlarger{\mathlarger{\sum}}_B \,\frac{\gamma_B}{2\pi^2} \bigintsss_0^{k_B} \sqrt{k^2 + m^{*2}_B(\bar{\sigma})}\,k^2 \, dk + \mathlarger{\mathlarger{\sum}}_l \, \frac{1}{\pi^2} \bigintsss_0^{k_l} \sqrt{k^2 + m^{2}_l(\bar{\sigma})}\, k^2 \, dk\\
&+& \frac{1}{2} m_{\sigma}^2\, \bar{\sigma}^2 + \frac{1}{2} m_{\omega}^2 \,\bar{\omega}^2 + \frac{1}{2} m_{\rho}^2 \,\bar{\rho}^2
+ \frac{1}{3} b_{\sigma} m_N (g_{\sigma N}\, \bar{\sigma})^3 + \frac{1}{4} c_{\sigma} (g_{\sigma N}\, \bar{\sigma})^4 \, .
\end{array}
\end{equation}

The saturation properties including the nuclear saturation density,
$n_0$, energy per nucleon, $E_0$, nuclear incompressibility, $K_0$,
effective nucleon mass, $m^*/m_N$, asymmetry energy, $J$, asymmetry
energy slope, $L_0$, and nucleon potential, $U_N$, and the parameters
of the GM1(L) model used in this work are listed {in Tables}
\ref{table:properties} and \ref{table:parametrizations}, respectively.
{{Besides the RMF, there are other approaches to construct the EoS that describes hadronic matter. One option is to calculate the hadronic EoS performing microscopic calculations through the Brueckner-Bethe-Goldstone many-body theory with two and three body nuclear
interactions~\cite{BBB1997A&A...328..274B}, which satisfies several requirements of a ``realistic'' EoS (adequate saturation point for symmetric nuclear matter, appropriates symmetry energy and incompressibility). Effective field theory with and without pions considering effective potentials to describe two-body low-energy observables~\cite{2018A&A...609A.128B} has also been used to describe nuclear
matter and NSs. Alternatively, the new microscopic EoS of dense asymmetric and $\beta-$stable nuclear matter at zero temperature presented in Ref.~\cite{Kievsky2018PhRvL.121g2701K} is another option, where two-body and three-body nuclear interactions derived in the context of chiral perturbation theory and Delta isobars were considered.}}

\begin{table}[H]
\centering
\captionsetup{width=1\linewidth}
\caption{Properties of nuclear matter at saturation density for the
    hadronic GM1(L) parametrization \cite{Spinella2017:thesis,Spinella:2018bdq}.}
\label{table:properties}
\begin{tabular}{cc}
\toprule
{\bf Saturation Property} & {\bf GM1L} \\
\midrule
$n_0$  (fm$^{-3}$)    & 0.153      \\
$E_0$  (MeV)          & $-16.30$    \\
$K_0$  (MeV)          & 300.0       \\
$m^*$/$m_N$             & 0.70        \\
$J$    (MeV)          & 32.5        \\
$L_0$  (MeV)          & 55.0        \\
$-U_N$ (MeV)        &65.5          \\
\bottomrule
\end{tabular}
\end{table}

\begin{table}[H]
\centering
  \caption{Parameters of GM1(L) that produce the properties of symmetric nuclear matter at saturation density given in Table~\ref{table:properties}.}
  \label{table:parametrizations}
\begin{tabular}{cc}
\toprule
{\bf Parameter} & {\bf  GM1L}\\
\midrule
$m_{\sigma}$  (GeV)    & 0.5500       \\
$m_{\omega}$  (GeV)          &0.7830    \\
$m_{\rho}$  (GeV)          & 0.7700      \\
$g_{\sigma N}$             & 9.5722      \\
$g_{\omega N}$            & 10.6180      \\
$g_{\rho N}$            & 8.9830      \\
$b_{\sigma}$         &0.0029           \\
$c_{\sigma}$         &$- 0.0011$         \\
$a_{\rho}$         &0.3898        \\
\bottomrule
\end{tabular}
\end{table}

\subsubsection{Phase Transition Formalism}
Depending on the value of the hadron-quark surface tension, $\sigma _{\rm HQ}$, the phase transition between hadronic and quark matter might be sharp (described via the Maxwell construction, which assumes that $\sigma _{\rm HQ}$ is infinitely large) or soft (that could be described via the Gibbs construction). On the other hand, ``Bulk'' Gibbs construction assumes that $\sigma _{\rm HQ}$ is zero while for intermediate values of $\sigma _{\rm HQ}$, the ``full'' Gibbs formalism must be used~\cite{2017PhRvC..96b5802W}.  A critical value for $\sigma _{\rm HQ}$ has been estimated to be $\sim$5--40\,${\rm MeV/fm}^2$ \citep{Alford:2001zr,Endo:2011em}. If the hadron-quark surface tension were grater than this value, a sharp phase transition would be favored. On the contrary, a mixed phase in which hadrons and quarks coexist may~occur.

In this work we present results using the Maxwell construction. In addition, we mimic mixing and percolation of the Gibbs construction using an interpolation function, $I_{\pm}(p)$, between the hadronic, $\epsilon_{\rm H}(p)$ and quark phases $\epsilon_{\rm Q}(p)$
\begin{equation} \label{interpolate}
I_{\pm} (p) = \frac{1}{2}\left[1 \pm \tanh \left(\frac{p-p_{\rm trans}}{10 {\rm {b}}\,p_{\rm trans}} \right) \right],
  \end{equation}
\noindent  where + ({$-$}) is used to characterize the function used for pressures greater (smaller) than the transition pressure, $p_{\rm trans}$, and $\rm{b}$ can be interpreted as a {\it mixing length} \cite{Macher:2004vw,2013PTEP.2013g3D01M,2013ApJ...764...12M,2015PPN,Alford:2017vca,Ranea-Sandoval:2018bgu}. {{The {\it mixing length} serves to quantify the extension of the mixed phase inside the compact object. Results obtained using this approach are qualitatively similar to those presented in Ref.~\cite{2017PhRvC..96b5802W} in which the authors include finite-size effects on the hadron-quark mixed phase in the context of compact objects. The extension of the mixed phase increases when both $\sigma _{\rm HQ}$ and ${\rm b}^{-1}$ decrease. This fact, although it is not clear (as there are several model-dependent phenomena), seems to indicate the existence of a relationship between these two quantities. Finding and analyzing this possible relationship is well beyond the scope of this work.}} 
Considering the mimicking of the mixed phase, the hybrid EoS reads
\begin{equation} \label{mixEOS}
\epsilon_{\rm MIX}(p) = \epsilon_{\rm H}(p)I_-(p)+ \epsilon_{\rm Q}(p)I_+(p).
  \end{equation}

\begin{figure}[H]
  \centering
  \includegraphics[width=0.4\textwidth]{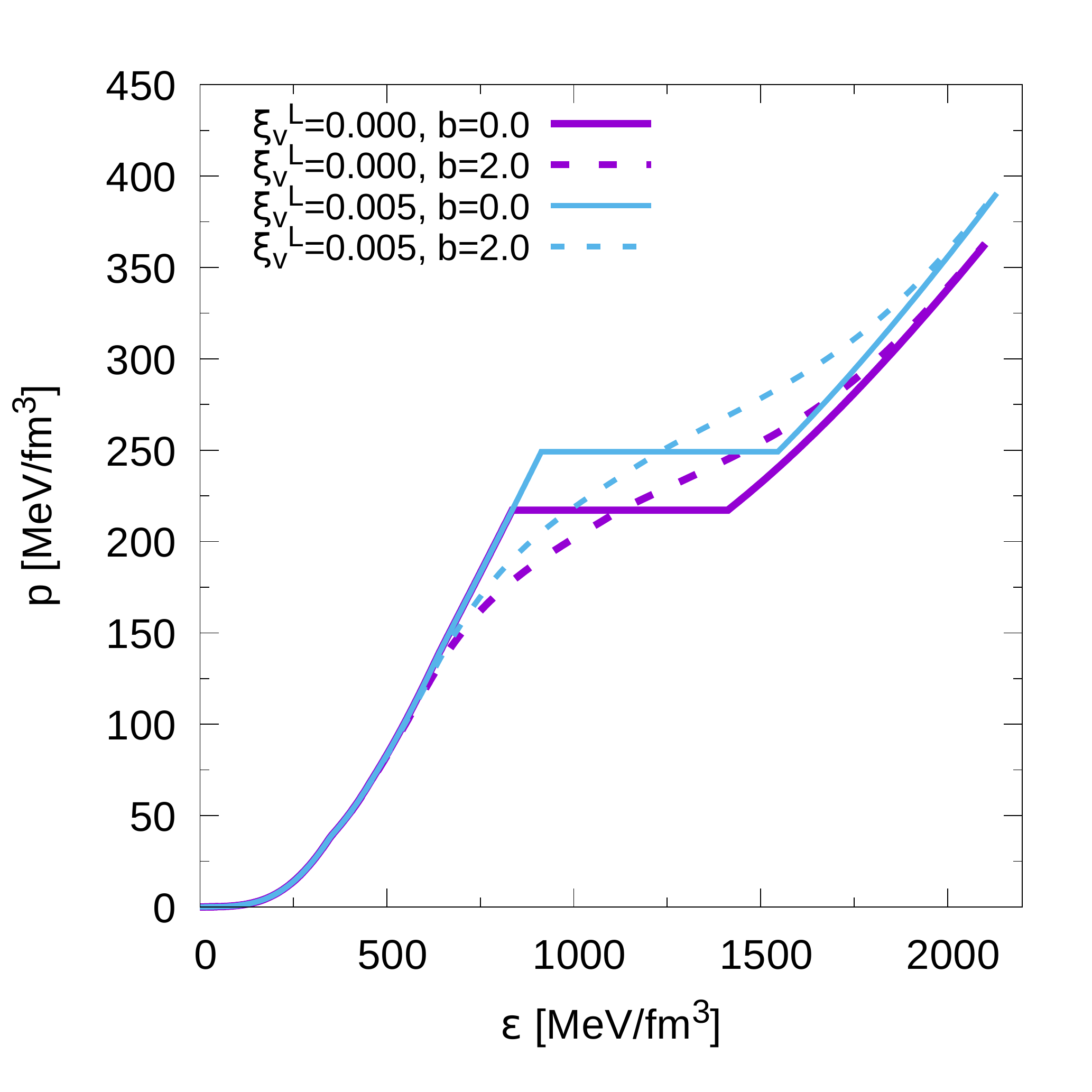}
  \includegraphics[width=0.4\textwidth]{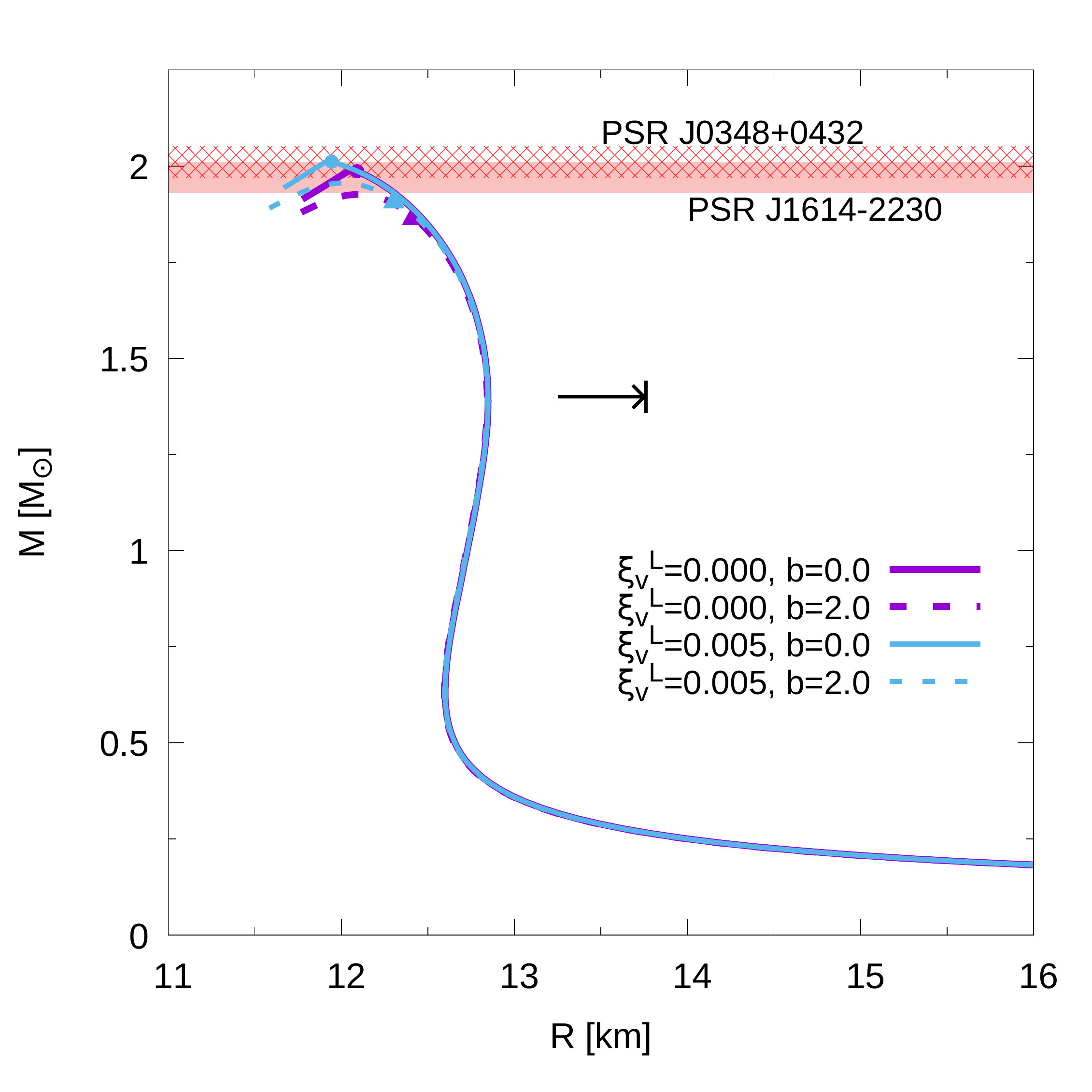}
  \caption{Hybrid EoSs ({\bf left}) and mass-radius (M-R) relationship ({\bf right}) for the cases corresponding to the local NJL EoSs. In the EoS curves, the constant pressure regions for the $b=0.0$ cases correspond to the transition phase. The $b=2.0$ cases does not have an abrupt transition so they have a mixed phase region. In the M-R curves, the rounded dot {{on the solid line}} indicates where the quark mater core appears {{and the triangle on the dashed curve indicates the first stellar configuration in which mixed phase is present in its inner core}}. After the peaks, towards smaller radii, the stars become unstable. The horizontal bars are the measured masses of the $2~M_\odot$ pulsars with their corresponding errors. The horizontal arrow marks the constraint calculated in Ref.~\cite{Fattoyev:2017jql} for GW170817, $R_{M=1.4 M_\odot}<13.76~{\rm km}$.
}
  \label{eos-mraio-local}
\end{figure}

\begin{figure}[H]
  \centering
  \includegraphics[width=0.45\textwidth]{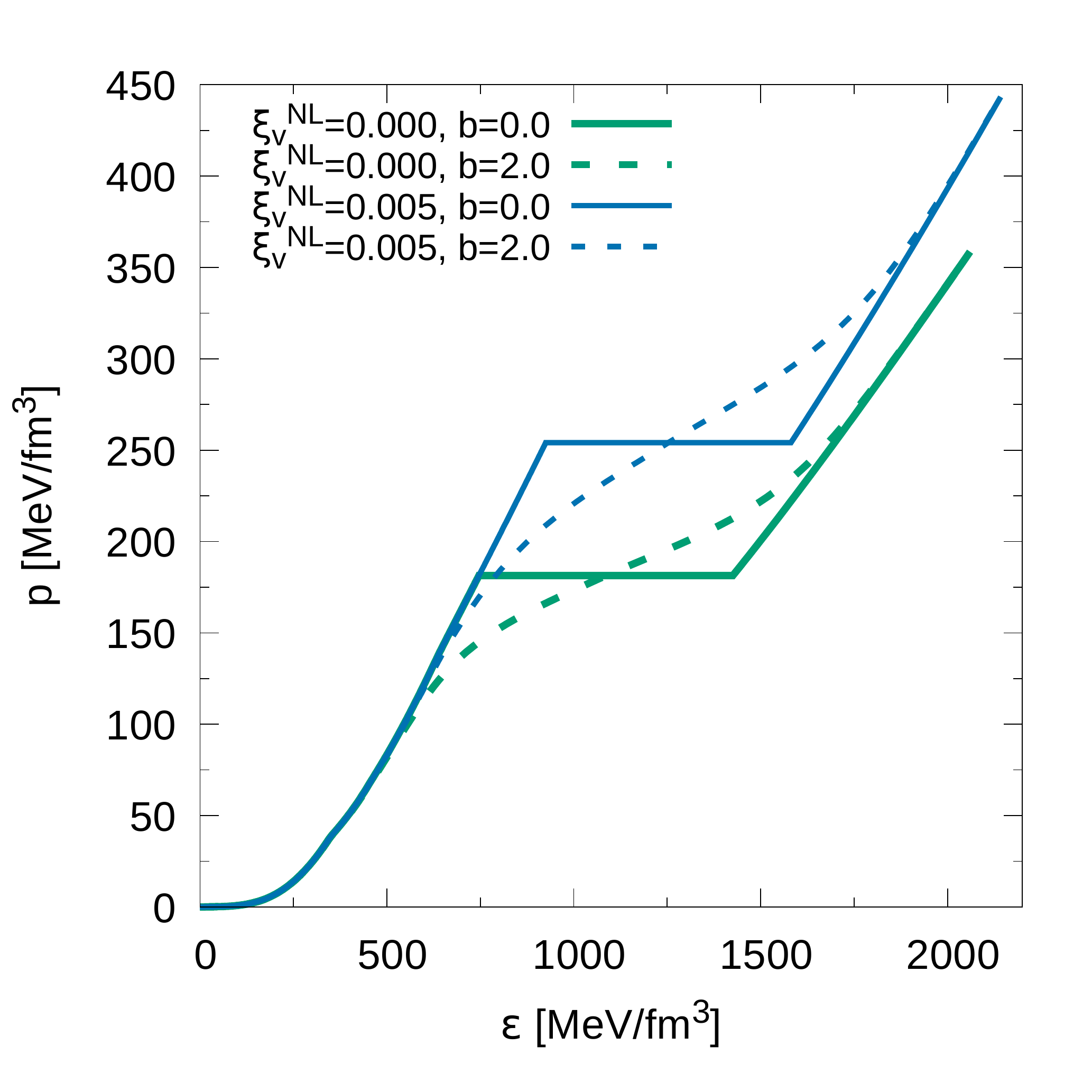}
  \includegraphics[width=0.45\textwidth]{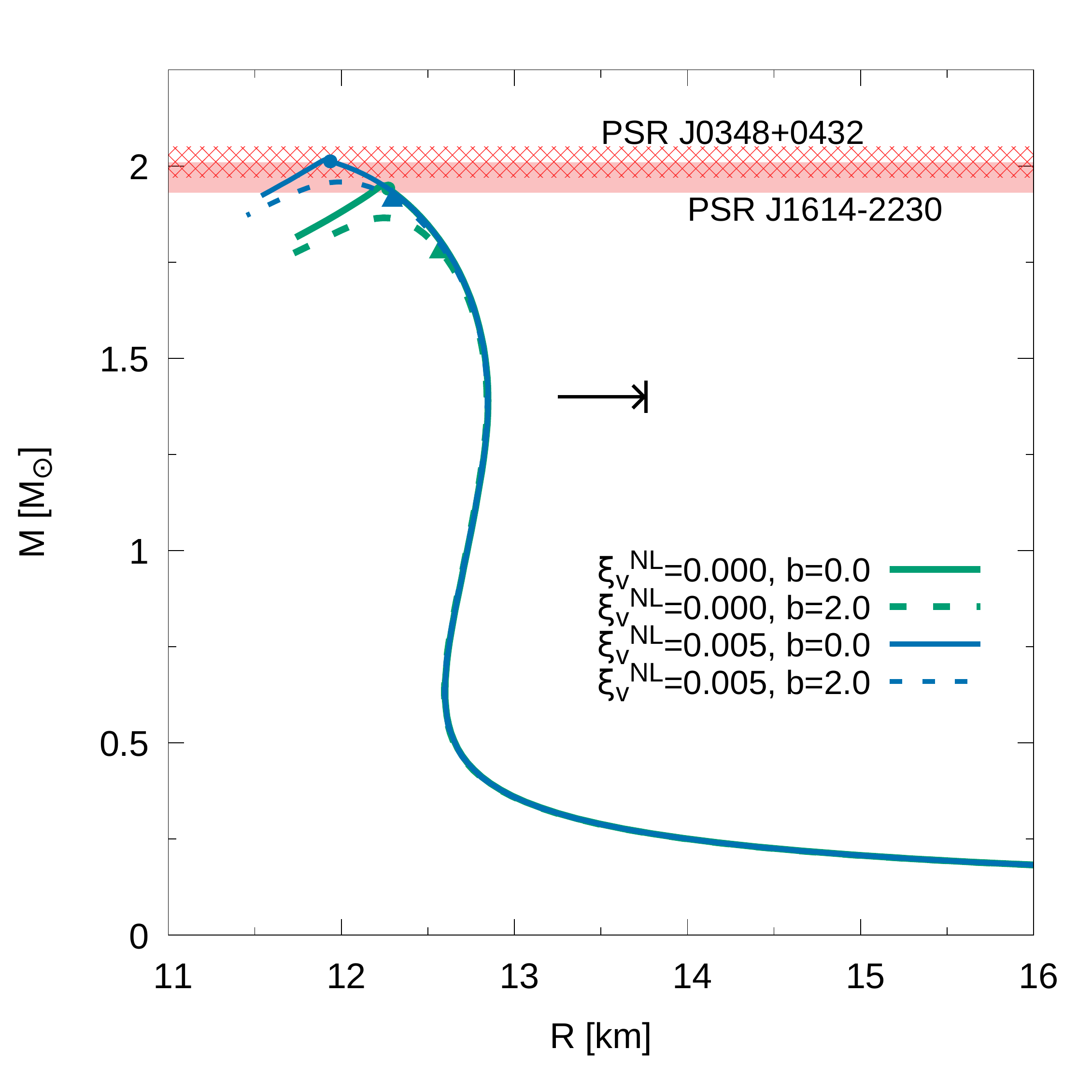}
  \caption{Same as Figure~\ref{eos-mraio-local} but for the non-local NJL.
}
  \label{eos-mraio-nolocal}
\end{figure}

The inclusion of the interpolating function presented in Equation~(\ref{interpolate}) allows us to qualitatively study and analyze how the nature of the phase transition alter the structure and oscillation frequencies of hybrid stars. To study potential observable effects on such quantities, we use two different mixing lengths: $\rm{b} = 0$ (Maxwell construction) and $\rm{b} = 2$. {{To calculate the onset of a mixed phase region, we assume that it appears at the pressure where the energy density obtained using Equation~(\ref{mixEOS}) differs from the pure hadronic EOS by a 5\%. Using a similar approach, we also estimate the appearance of a pure quark core.}} {{Alternative methods to describe the hadron-quark phase transition using different interpolation prescriptions have been presented and discussed in Refs.~\cite{2018Univ....4...94A,2018arXiv181211889M}. Our results are in general agreement with those obtained by such approximations.}}
{In left panel of Figure} \ref{eos-mraio-local} we present the hybrid EoS constructed using the local NJL quark matter model. With solid lines we show hybrid EoS where the phase transition is constructed using Maxwell criteria for two different values of the vector coupling constant, $\xi_{\rm v}^{\rm L}$. The dashed lines correspond to the hybrid EoS with a mixed phase mimicked using the interpolating function (\ref{interpolate}). In the left panel of Figure~\ref{eos-mraio-nolocal} we present the same but for hybrid EoS constructed using non-local NJL quark model with vector interaction, $\xi_{\rm v}^{\rm NL}$.

The structure of NSs strongly depends on the EoS used to describe matter inside the star. Given a particular EoS, the TOV equations allow to obtain the corresponding family of stationary stellar models. These theoretical curves can be used to discard models that are unable to reproduce astronomical observations~\cite{Ozel:2016oaf}.

In right panel of Figure~\ref{eos-mraio-local} we present the M-R relationships obtained using the local NJL quark matter model. In right panels of Figures~\ref{eos-mraio-local} and \ref{eos-mraio-nolocal} we present the M-R relationship obtained using local and non-local NJL models, respectively. The lines colors and styles are the same as used in the corresponding left panels. In both figures the horizontal bars are the measured masses and error bars of pulsars PSR J0348+0432 and PSR J1614-2230. The horizontal arrow represents $R_{1.4 M_\odot}<13.76$ km the constraint calculated using data of GW170817~\cite{Fattoyev:2017jql}. {{For the ${\rm b} = 2$ cases, we have computed the appearance of a mixed phase obtaining an extended branch of hybrid stellar configurations. In all the considered cases these branches spans over a range $\Delta R$$\sim$$0.3$ km and $\Delta M$$\sim$$0.04~M_\odot$. For these cases, none of the stable stellar configurations present a pure quark matter core. This result is similar to the one obtained using Gibbs formalism to treat the hadron-quark phase transition~\cite{Orsaria:2013,Orsaria:2014}.}}

\subsection{Stellar Oscillations}
There are two distinct families of fluid perturbation modes: {\it odd} (or axial), which produce toroidal deformations and {\it even} (or polar) which produce spheroidal ones. From an observational point of view, the most promising gravitational waves (GWs) are those related to the fundamental ($f$), the pressure ($p$) and the gravity ($g$) fluid modes. The $g$-modes are particularly interesting as they appear in cold, non-rotating NSs  only if a discontinuous EoS is needed to describe matter in their interior~\cite{Finn:1987}. For this reason, a detection of such modes would be a clear indicator of the existence of a sharp hadron-quark phase transition in the inner core of such compact object.

When the metric perturbations are negligible, the equations that govern linear perturbations are greatly simplified. This idea, first applied in Newtonian theory~\cite{cowling:1941}, was then extended to study relativistic scenarios~\cite{McDermott:1983}. This simplification is called relativistic Cowling's approximation.

Isolated NSs are (in almost all situations) very accurately described using spherically symmetric metrics. In such cases, the equations needed to solve in order to find these frequencies can be written~as
\begin{eqnarray} \label{eq:modes}
  \frac{{\rm d}W (r)}{{\rm d}r} &=& \frac{{\rm d} \epsilon}{{\rm d}P} \left[\omega ^2 r^2 {\rm e}^{\Lambda (r) - 2 \Phi (r)} V(r) + \frac{{\rm d}\Phi (r)}{{\rm d}r} W(r)\right] - \ell (\ell + 1){\rm e}^{\Lambda (r) } V(r), \nonumber \\
  \frac{{\rm d}V(r)}{{\rm d}r} &=& 2 \frac{{\rm d}\Phi (r)}{{\rm d}r} V(r) - \frac{1}{r^2}{\rm e}^{\Lambda (r) } W(r) .
  \end{eqnarray}

\noindent The functions $\Phi (r)$ and $\Lambda (r)$ characterize the background spacetime and the functions $V(r)$ and $W(r)$, along with the frequency $\omega$, characterize the Lagrangian perturbation vector of the fluid,
{
\begin{equation} \label{pert}
\zeta ^i = \left(e^{-\Lambda (r)}W(r), -V(r)\partial _\theta, -V(r) \sin ^{-2} \theta \partial _\phi \right)r ^{-2} Y_{\ell m}(\theta , \phi),
\end{equation}
}
\noindent where $Y_{\ell m}(\theta , \phi)$ is the $\ell m$-spherical harmonic~\cite{sotani:2011}. For details related to the physically correct boundary conditions and the numerical procedure used to solve Equations (\ref{eq:modes}), see Ref.~\cite{Ranea-Sandoval:2018bgu}.

The frequencies obtained, for quadrupolar perturbations, $\ell = 2$, using the relativistic Cowling's approximation are in general in agreement with those obtained using linearized general relativity showing both qualitatively and quantitative good results for the $p$ and $g$-modes and only a qualitatively good results for the $f$-modes with errors $\sim$15--20\% (for details see,~\cite{Yoshida:1997,chirenti:2015,Ranea-Sandoval:2018bgu}, and references therein). We have calculated the frequencies of the fundamental, $f$, and first pressure, $p_1$-modes for each of the EoSs presented in Section \ref{hyb} using both the sharp and mixed phase formalism to construct the phase transition.  For the discontinuous EoSs we have also calculated the gravity, $g$, modes associated with the energy gap, $\Delta \epsilon$, between the hadronic and quark phases. 

\begin{figure}[H]
  \centering
  \includegraphics[width=0.35\textwidth,angle=-90]{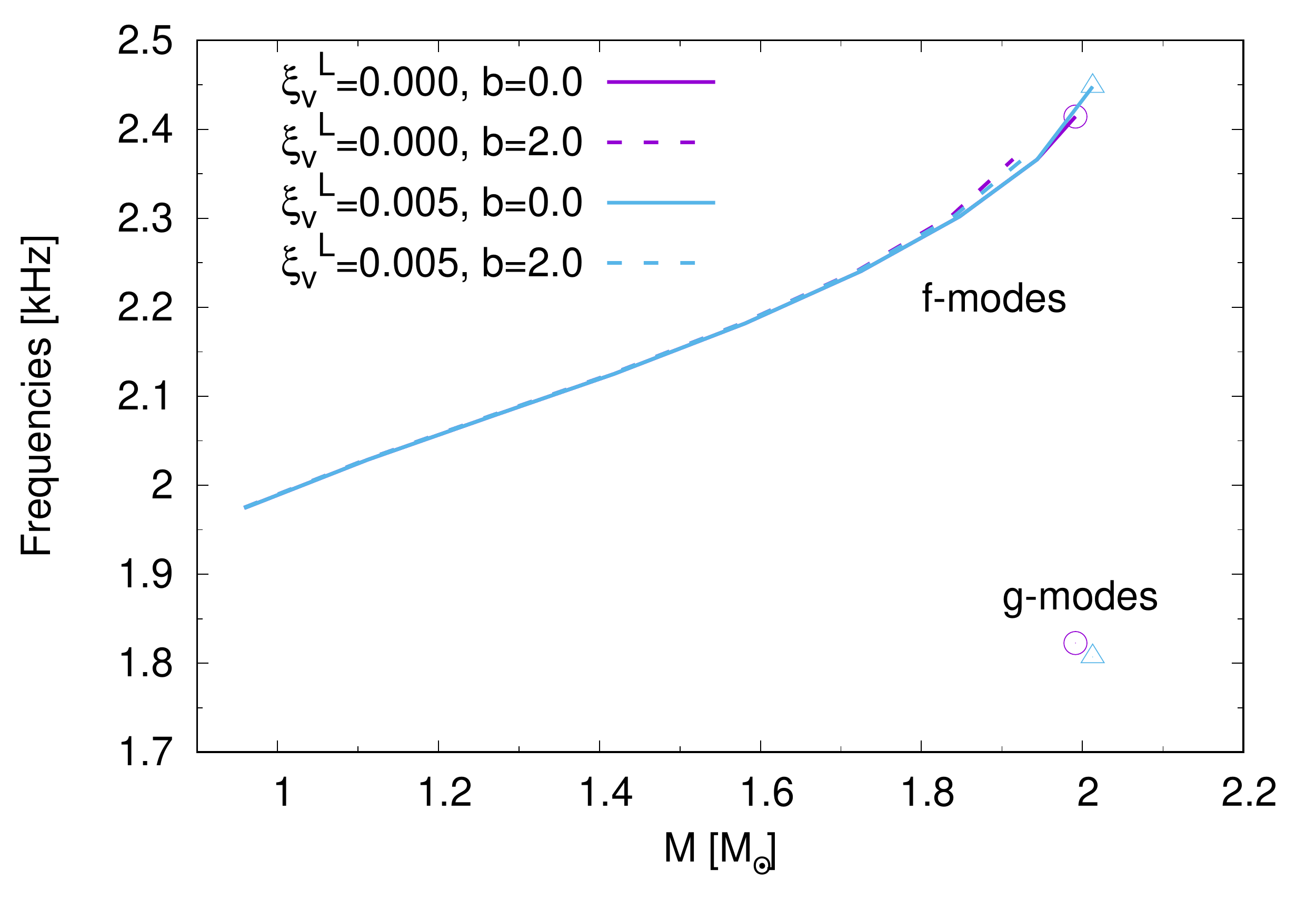}\includegraphics[width=0.35\textwidth,angle=-90]{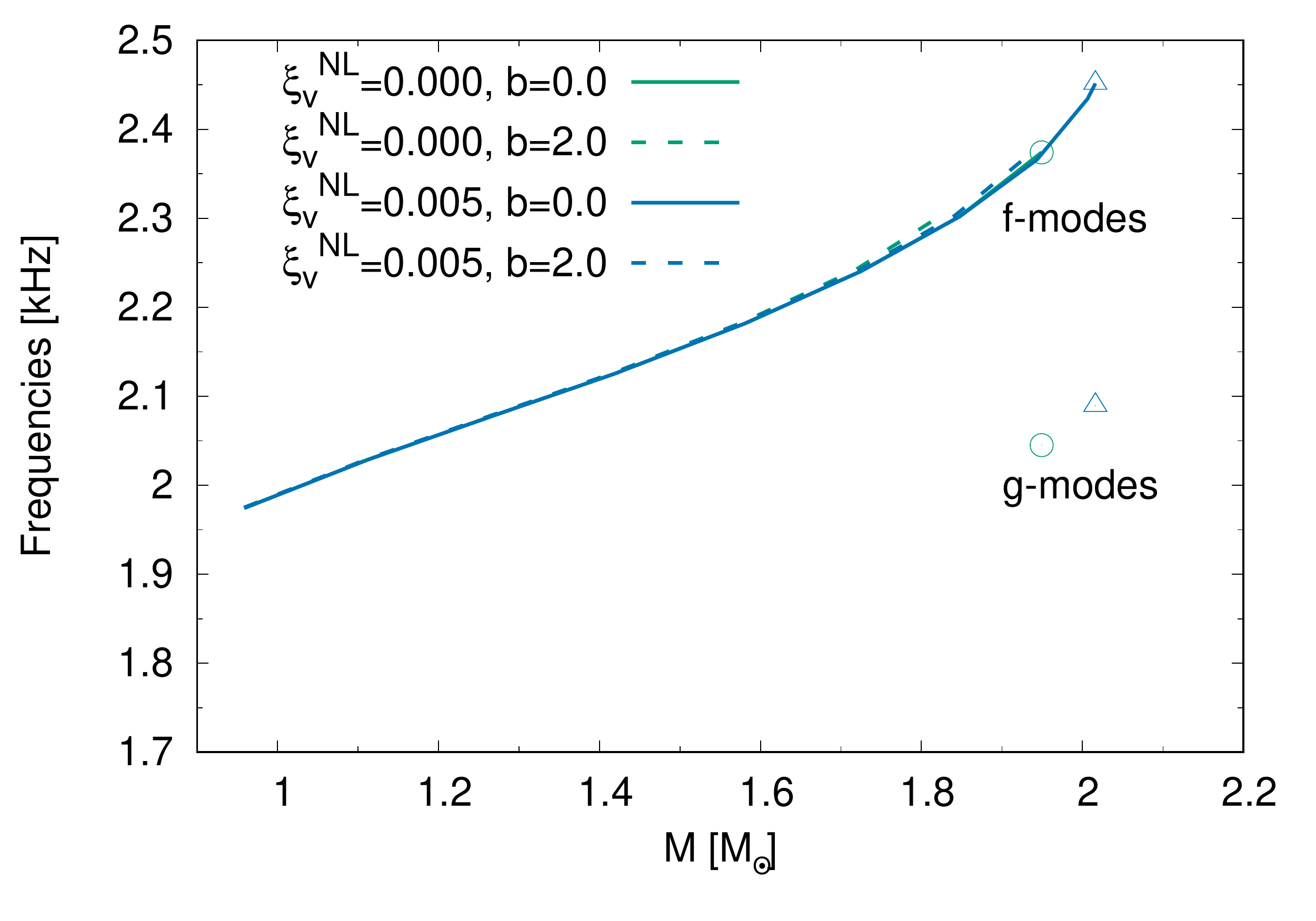}
  \caption{Frequencies of the $f$ and $g$-modes (if present) for the hybrid EoSs constructed with the local model are presented in {\bf left} panel while those obtained using the non-local model are presented in the {\bf right} one. With solid lines we present models with sharp phase transition and with dashed lines those with a mixed phase. With circle and triangle the frequency of the quark hybrid stars.
}
  \label{g-and-f}
\end{figure}

\begin{figure}[H]
  \centering
  \includegraphics[width=0.35\textwidth,angle=-90]{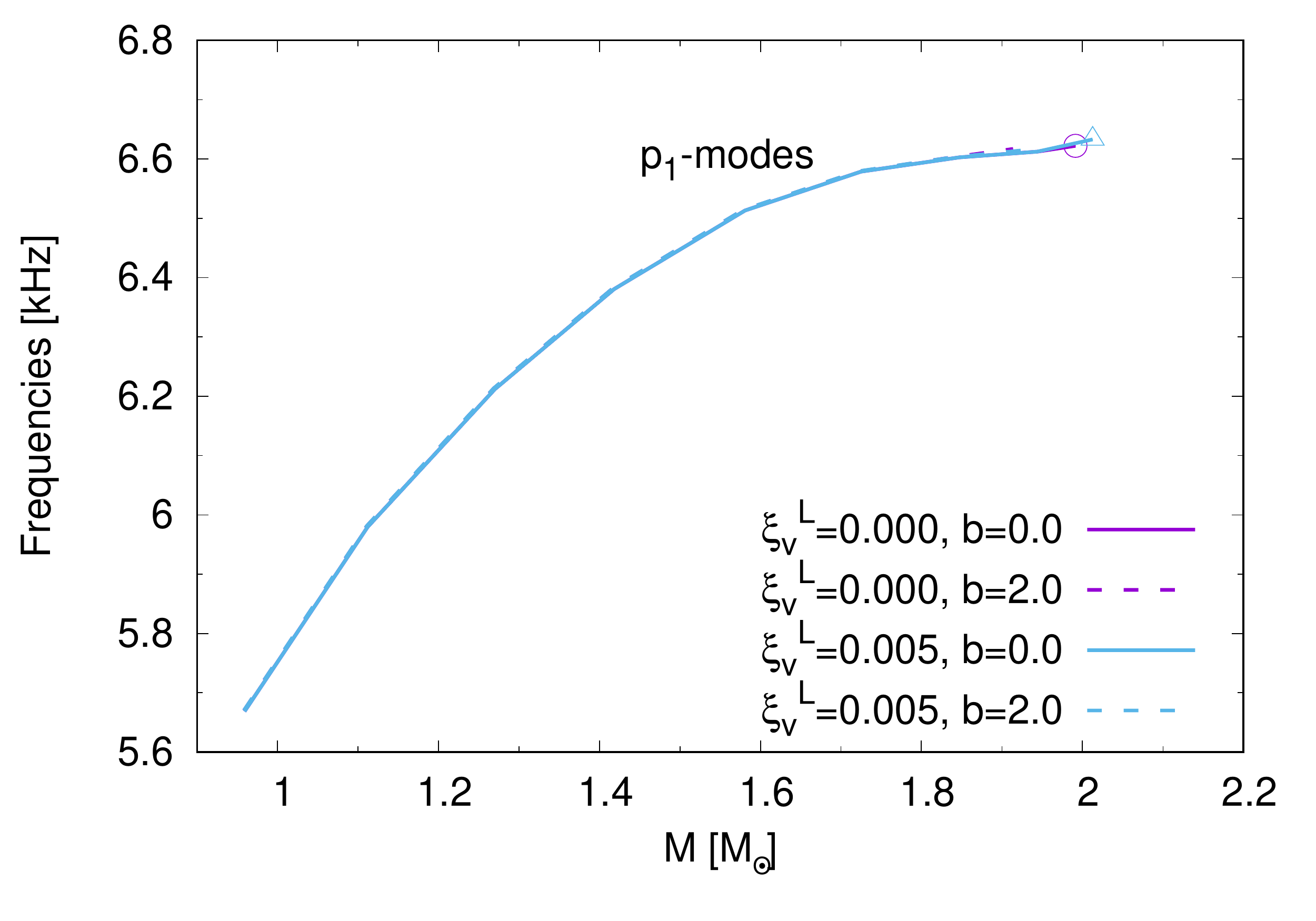}\includegraphics[width=0.35\textwidth,angle=-90]{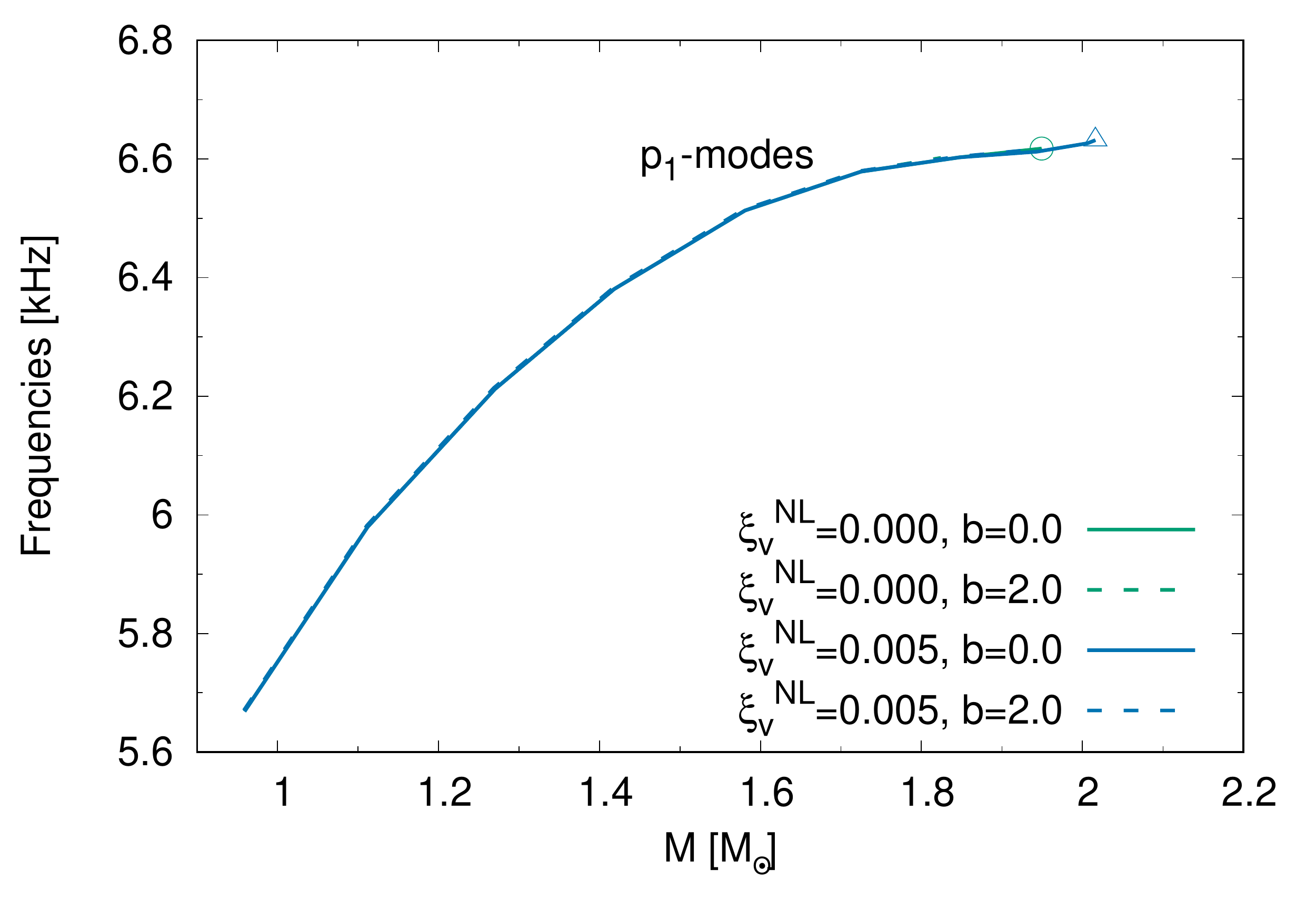}
  \caption{Frequencies of the $p_1$-modes as a function of the compact object. In the {\bf left} panel we show those corresponding to hybrid EoSs constructed with the local model. In the {\bf right} panel we present those obtained using non-local model. With solid lines we present models with sharp phase transition and with dashed lines those with a mixed phase. With circle and triangle the frequency of the quark hybrid stars.
}
  \label{p1}
\end{figure}

The results for the $f$ and $g$-modes are presented in the left (right) {panel of Figure} \ref{g-and-f} for the hybrid EoSs constructed using local (non-local) NJL model. For the local (non-local) NJL model, the results of the first pressure are presented in the left (right) {panel of Figure} \ref{p1}.
The general result is that a detection of $f$ and $p_1$ modes is not enough to determine nor the quark matter model used nor the nature of the phase transition. Only a simultaneous detection of a $g$ and $f$-mode would indicate the occurrence of a sharp hadron-quark phase transition in the inner core of NSs. Similar results where obtained in Ref.~\cite{Ranea-Sandoval:2018bgu}.

\unskip

\begin{figure}[H]
  \centering
  \includegraphics[width=0.55\textwidth]{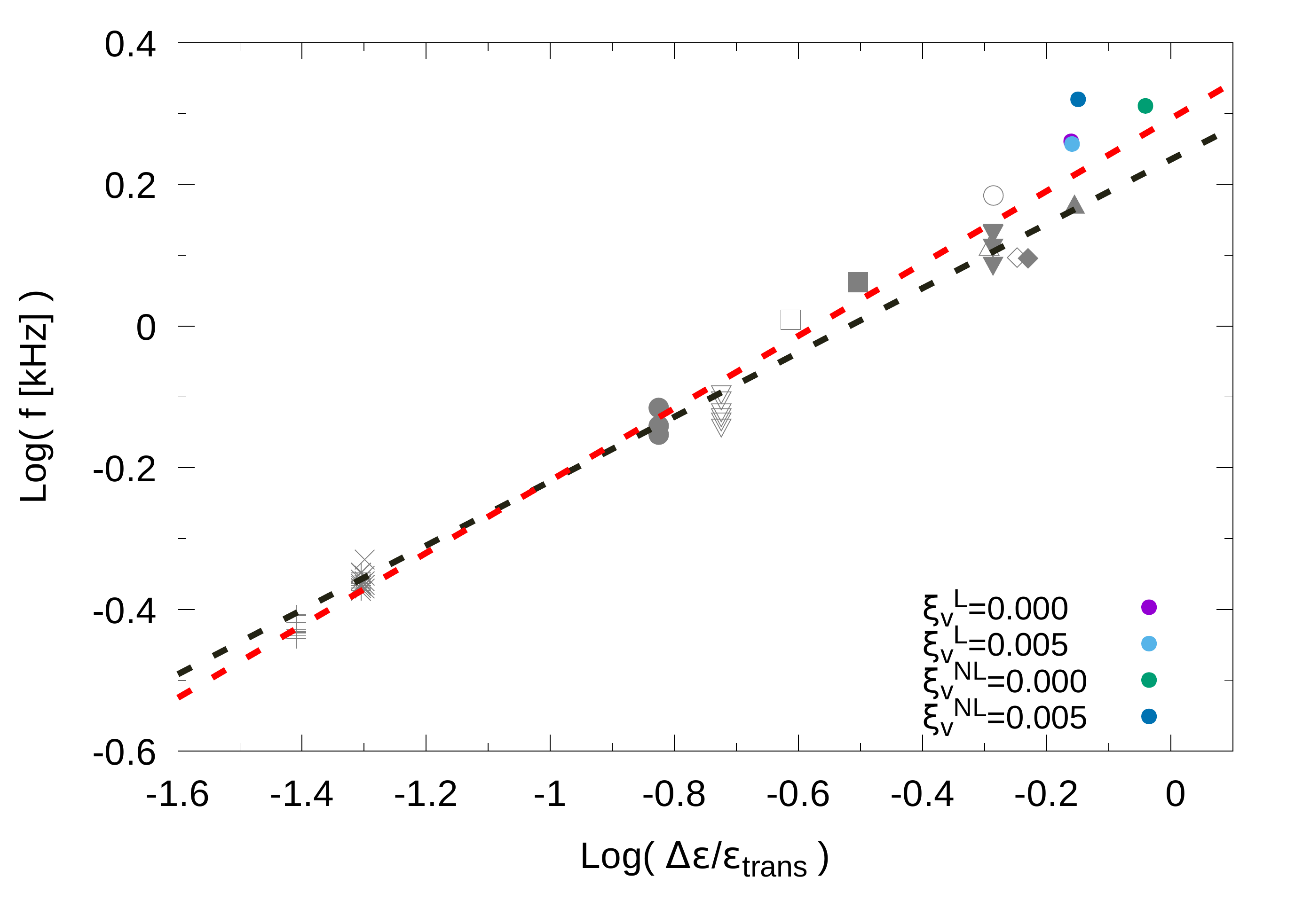}
  \caption{Represented with dots, decimal logarithm of the frequencies of the $g$-modes obtained in this work as a function of the logarithm of the $\Delta\epsilon/\epsilon_{\rm trans}$. With red dashed line we present the revised {\it universal} relationship between these quantities. Using gray scale the results presented in~\cite{Ranea-Sandoval:2018bgu}.
}
  \label{universal_g}
\end{figure}

In addition, in Figure~\ref{universal_g}, we present a revisited universal relationship obtained after the inclusion of the additional frequencies associated with the EoSs used in this work. The universal relationship proposed in Ref.~\cite{Ranea-Sandoval:2018bgu} between  $x_{\rm CSS} \equiv \Delta \epsilon / \epsilon _{\rm trans}$, one of parameters of the Constant Speed of Sound (CSS) parametrization for quark matter~\cite{CSS-original,Ranea:2016}, and the frequency of the $g$-modes. The original fitting function is $y_{\rm fit}^{\rm old} = c_1^{\rm old}  \log (x_{\rm CSS}) + c_2^{\rm old}$ with $c_1^{\rm old} = 0.454 \pm 0.031$ and $c_2^{\rm old} = 0.235 \pm 0.023$. The root mean square of the residuals of the fit is $R_{\rm rms}^{\rm old} = 0.048$ showing that the proposed correlation is strong. According the additional frequencies, the relationship suffers only small quantitative changes:  $y_{\rm fit}^{\rm new} = c_1^{\rm new}  \log (x_{\rm CSS}) + c_2^{\rm new}$ with $c_1^{\rm new} = 0.511 \pm 0.036$ and $c_2^{\rm new} = 0.2935 \pm 0.024$. The root mean square of the residuals of the fit is $R_{\rm rms}^{\rm new} = 0.065$. The tight fit is still present, making this {\it universal} relationship a potentially powerful tool to perform NS Asteroseismology. 

\unskip

\section{Results and Discussion}
\label{res}

In this work we have constructed hybrid EoSs using the GM1(L) parametrization to describe the hadronic phase and local and non-local NJL models including vector interactions to model the quark matter phase. The phase transition is constructed both using the Maxwell construction considering a sharp phase transition and an interpolation function that mimics mixed phase effects for the Gibbs construction. The main effect of the mixed phase is to reduce the maximum mass of the stationary family of stellar configurations. Models in which we consider sharp phase transition show stellar configurations with pure quark matter in their cores, but its appearance inmediately destabilizes the star as it has been shown in several previous works (see, for example, Refs.~\cite{2015PhRvD..92h3002A,Ranea:2016,Ranea:2017}). {{It is worth to mention that in this work we have not included color superconductivity effects that are expected to be relevant at large densities (see, for example, Refs.~\cite{PhysRevD.79.096008, 2008RvMP...80.1455A} and references therein). Color superconductivity is known to lower the hadron-quark transition pressure giving rise to longer branches of quark hybrid stars (see for example, Ref.~\cite{Ranea:2017} and references therein). }}

In addition, we have studied the QCD phase diagram by compute the phase transition curve for local and non-local NJL models including Polyakov loop. The non-local NJL parametrization used in this work produces a first order phase transition and has no critical end point. This result do not agree with LQCD results. On the contrary, the local model do present a crossover as it is expected, the CEP is located at $T_{CEP} = 103.2$ MeV and $\mu_{CEP} = 953.4$ MeV. We have considered $T_0 = 195$ MeV as the critical temperature of the Polyakov loop effective potential. A different anzat of the Polyakov loop potential~\cite{Fukushima:2008wg,Fukushima:2008b} as well as other choices of $T_0$ could improve the obtained results regarding the critical temperature at zero chemical potential ($T_c = 164.9$ MeV for the non-local model and $T_c = 198.5$ MeV for the local model).

The models developed in this work have been compared to Lattice QCD data calculations and experimental chemical freeze-out results, showing better agreement when non-locality is taken into account. In general, several studies using non-local NJL models have shown a more realistic and consistent description of the strongly interacting matter at extreme conditions of temperature and/or density  than the local NJL models (see for example Refs.~\cite{Contrera:2010, PhysRevD.84.056010, PhysRevD.96.114012} and references therein). However, the parametrization of the non-local model used in this work leads to results that are not compatible with the predictions of Lattice QCD~\cite{Ratti_2018} regarding the QCD phase diagram. The new parametrization used in Ref.~\cite{jpg2019} solves completely these discrepancies, as it is shown in Figure~\ref{qcd-diag}. 

We have also presented, for quadrupolar perturbations, $\ell = 2$, the frequency modes ($f$, $p_1$ and $g$, if they exists) for the given EoSs. The detection of such modes is not a powerful tool to discriminate between different NJL models since the differences between the frequencies of the $f$ and $p_1$-modes for a given stellar mass is less that $\sim$1\% for the most massive stars. Moreover, unless a $g$-mode is detected, gravitational waves would not be enough to determine the nature of the phase transition since the effects of the appearance of a mixed phase produce no observable effects on such frequencies. {{It is expected,  as presented in Ref.~\cite{Ranea-Sandoval:2018bgu}, the frequencies of the calculated $g$-modes become higher for stars with a color superconducting quark matter phase in their cores.}} Additionally, we have  obtained slight modifications to the fit parameters of the {\it universal} relationship between the frequency of the $g$-modes and the  $x_{\rm CSS}$ parameter obtained in~\cite{Ranea-Sandoval:2018bgu}. This relationship, and other existing ones (see, for example,~\cite{AndersonKokkotas:1998,GWfromNS}) could shed some light into the problem of determining NSs parameters (such as mass, radius and internal composition) from GWs observations. 

The analysis of neutron stars with the presence of deconfined quark mater in their cores, shows that for the NJL models considered in this work a quark matter core is only possible for the last stable star in the M-R diagram. The presence of diquarks can give longer stable hybrid branches~\cite{Ranea:2017}.

\vspace{6pt}

\authorcontributions{I.F.R.-S. and M.G.O. supervised the project, wrote most of the article with input from the other authors and performed great part of the analysis of the obtained results. G.M. and D.C. performed calculations related to the phase diagrams of NJL models and helped with the analysis of the results. M.M. calculated the hybrid EOSs at zero temperature and obtained the stationary configurations and produced the corresponding figures. G.A.C. wrote the Subsubsection in which the GM1(L) hadronic model is described and parts of Subsection in which the Non-local SU(3) model is presented. Moreover, he produced some of the figures included in the text. O.M.G. calculated oscillation frequencies within the Cowling approximation and produced the corresponding figures. All the authors discussed the results and commented on the final version of the~manuscript.
}

\funding{This research was funded by CONICET grants number PIP-0714 and PIP-0436 and Universidad Nacional de La Plata under grants number G140, G157, X824 and G144. O.M.G. is also supported by the PICT 2016-0053 from ANPCyT, Argentina.}

\acknowledgments{The authors want to thank Tomohiro Inagaki, Emilio Elizalde and Dalia Su for inviting us to publish our work in the special issue of the journal Symmetry ``Nambu-Jona-Lasinio model and its applications''. We would also like to thank the anonymous referees for their comments and suggestions that help improve the original manuscript. O.M.G. also acknowledges the hosting as invited researcher from IA-PUC. G.A.C. also acknowledges the hosting at the SDSU, USA, within the CONICET-NSF joint research project titled {\it Structure and properties on neutron star cores}.}

\conflictsofinterest{There are no conflicts of interest regarding this research work.}


\reftitle{References}

\newcommand{\apj}{Astrophys. J.}
\newcommand{\prd}{Phys. Rev. D}
\newcommand{\prc}{Phys. Rev. C}
\newcommand{\apjl}{Astrophys. J. Lett.}
\newcommand{\mnras}{Mon. Not. R. Astron. Soc.}
\newcommand{\aap}{Astron. Astrophys.}
\newcommand{\jcap}{J. Cosmol. Astropart. Phys.}

\end{document}